\documentclass[11pt]{article}

\usepackage{graphicx}
\usepackage{epstopdf}
\usepackage{amsmath}
\usepackage{amssymb}
\usepackage{amsfonts}
\usepackage{amsthm}
\usepackage[usenames]{color}
\usepackage{array}

\usepackage[
      colorlinks=true,
      linkcolor=blue,
      urlcolor=blue,
      filecolor=blue,
      citecolor=red,
      pdfstartview=FitV,
      pdftitle={},
      pdfauthor={},
      pdfsubject={},
      pdfkeywords={},
      pdfpagemode=None,
      bookmarksopen=true
]{hyperref}

\usepackage{epsfig}

\usepackage{hyperref}

\renewcommand{\(}{\left(}

\newcommand{\bea}{\begin{eqnarray}}
\newcommand{\eea}{\end{eqnarray}}
\newcommand{\ba}{\begin{array}}
\newcommand{\ea}{\end{array}}
\newcommand{\ee}{\end{equation}}

\begin{document}

\begin{flushright}
\texttt{\today}
\end{flushright}

\begin{centering}

\vspace{2cm}

\textbf{\Large{
 Flat-space Holography and Correlators of \\ Robinson-Trautman Stress tensor  }}

  \vspace{0.8cm}

  {\large   Reza Fareghbal, Isa Mohammadi }

  \vspace{0.5cm}

\begin{minipage}{.9\textwidth}\small
\begin{center}

{\it  Department of Physics, 
Shahid Beheshti University, 
G.C., Evin, Tehran 19839, Iran.  }\\

  \vspace{0.5cm}
{\tt  r$\_$fareghbal@sbu.ac.ir, isa.physic2016@gmail.com}
\\ $ \, $ \\

\end{center}
\end{minipage}


\begin{abstract}
  We propose a quasi-local stress tensor for the four-dimensional  asymptotically flat Robinson-Trautman geometries by taking the flat-space limit from the corresponding  asymptotically AdS solutions. This stress tensor results in the correct charges of the generators of BMS symmetry if we define  conformal infinity by an anisotropic scaling of the metric components. Using flat-space holography this stress tensor is related to expectation values of the stress tensor in a dual field theory called  BMS-invariant field theory (BMSFT). We also  calculate the two and three point functions of the proposed stress tensor. 
\end{abstract}

\end{centering}

\newpage



\section{Introduction}
  According to the proposal of \cite{Bagchi:2010zz, Bagchi:2012cy}, the holographic dual of asymptotically flat spacetimes in (d+1) dimensions are $d$-dimensional field theories which are invariant under BMS symmetry (BMSFT). On the gravity side, BMS symmetry is the asymptotic symmetry of the asymptotically flat spacetimes at  null infinity \cite{BMS, Ashtekar:1996cd, Barnich:2006av, Barnich:4dBMS, aspects}. Similar to the AdS/CFT correspondence, the asymptotic symmetry of the gravity theory in the bulk is the same as the exact symmetry of the filed theory living on the boundary. BMS symmetry in three and four dimensions is infinite-dimensional.  Hence, similar to the two-dimensional conformal field theories (CFT$_2$),   one would expect some universal properties for the dual  BMSFT$_2$ and BMSFT$_3$. These properties can be studied by holographic methods in the context of Flat/BMSFT correspondence.
  
 Taking the flat-space limit (zero cosmological constant limit) from the asymptotically  AdS spacetimes, written in the suitable coordinates, results in the asymptotically flat spacetimes. It is plausible  to find some aspects of BMSFTs by starting from AdS/CFT correspondance and taking the flat space limit.  Most of the previous works have been concentrated on BMSFT$_2$ (see \cite{Prohazka:2017equ,Prohazka:2017lqb} for a complete list of related works). However BMSFT$_3$ is also interesting due to its holographic connection with  four-dimensional spacetimes. In this paper we focus on the latter case.
 
  Our approach in this paper is similar to  \cite{ Fareghbal:2013ifa,Fareghbal.Hosseni,Fareghbal:2016hqr,   Fareghbal.A.O}. We start from some asymptotically flat solutions and find the asymptotically AdS spacetimes whose flat-space limit leads to the original  asymptotically flat metric. Then, we use the dictionary of AdS/CFT to find the quasi-local stress tensor of the corresponding asymptotically AdS spacetimes \cite{Brown.AdS}. We take the flat-space limit of the stress tensor components and propose the quasi-local stress tensor of the asymptotically flat spacetimes. Using stress tensor components and applying Brown and York's method \cite{Brown.qe}, we calculate the corresponding charges of symmetry generators.  Our results are exactly the same as the charges  derived by covariant phase space method \cite{Barnich:2001jy, Wald.L, Wald.Z}. 
  
  The stress tensor components are given in terms of the parameters of the gravity solution. Variation of these parameters with respect to the symmetry generators is used to derive  variation of the stress tensor components. We can use them on the boundary side and impose  invariance of the stress-tensor correlators under the action of the global part of the BMS symmetry to derive the universal forms of  BMSFT$_3$ stress tensor one, two and three point functions. The final correlators are consistent with those of \cite{Bagchi:2016bcd}, where the results are obtained  by using the highest-weight representation of the BMS$ _4 $ algebra.
  
  We have used the above approach for  BMSFT$_2$ in  papers  \cite{ Fareghbal:2013ifa,Fareghbal.Hosseni,   Fareghbal.A.O} and for the quasi-local stress tensor of the Kerr black hole in \cite{Fareghbal:2016hqr}. In this paper we apply our method for a class of four-dimensional asymptotically flat spacetimes known as Robinson-Trautman (RT) solutions \cite{Robinson:1960zzb} (see also  \cite{RT} ). Schwarzschild black hole and non-rotating  time-dependent solutions are in this category. RT solutions with the negative cosmological constant have been studied in the context of AdS/CFT correspondence \cite{deFreitas:2014lia}-\cite{Ciambelli:2017wou}.  Although they are not invariant under the action of a generic BMS transformation, but it is not difficult to check that these types of solutions, with the zero cosmological constant,  satisfy the BMS boundary conditions  introduced in \cite{Barnich:4dBMS, aspects}. It is  straightforward to write down the corresponding asymptotically AdS geometry of   an asymptotically flat  RT type solution. This property makes it easy to holographically  study the asymptotically flat RT solution by starting from the AdS/CFT correspondence and taking the flat-space limit.
  
  The quasi-local stress tensor of the asymptotically AdS RT spacetimes are given by the dictionary of AdS/CFT correspondence. Although, taking the flat-space limit from the metric components is straightforward but stress tensor components are either zero or singular after taking the naive flat-space  limit. Hence, similar to  \cite{ Fareghbal:2013ifa,Fareghbal.Hosseni,Fareghbal:2016hqr,  Fareghbal.A.O}, we first multiply appropriate powers of AdS radius and then take the flat-space limit. In order to show that our proposed stress tensor is correct, we shall use it to compute the conserved charges of symmetry generators. To do so, we use the Brown and York's method \cite{Brown.qe} which needs an integration over a  surface. The well-known definition of conformal infinity for the four-dimensional  asymptotically flat spacetimes yields a two-dimensional surface. Apparently, BMSFT$_3$ cannot live on these surfaces. To solve the problem we define spacetimes where  BMSFT  coulde live on it by using an anisotropic scaling  of the asymptotically flat metric.  Currently we do not know a clear justification of this proposal but it should be   similar to the non-relativistic holography  discussed in \cite{Horava:2009vy}. In our previous works \cite{ Fareghbal:2013ifa,Fareghbal.Hosseni,Fareghbal:2016hqr,   Fareghbal.A.O}, we have successfully used this idea to compute charges by using proposed flat-space stress tensor. Our observation shows that if we  find an asymptotically flat spacetimes by taking flat-space limit from an asymptotically AdS spacetimes, the dual BMSFT   lives on a spacetime which has the same metric as the spacetime which CFT lives on it but the AdS radius should be replaced by proper powers of Newton's constant.   
  
 This paper is organized as follows: in section two, we briefly review Flat/BMSFT correspondence and Robinson-Trautman geometries. In section three we propose the quasi-local stress tensor of asymptotically flat Robinson-Trautman geometries and use it to compute the corresponding charges of the BMS symmetry. In section four we calculate the correlators of BMSFT$_4$ stress tensor. Last section is devoted to conclusions. 
  
 \section{Flat-space holography and Robinson-Trautman geometry}
Asymptotic symmetry of the asymptotically flat spacetimes at null infinity  in three and four dimensions is infinite dimensional \cite{Barnich:2006av,Barnich:4dBMS, aspects}. These symmetries which are the extensions of Poincare symmetry are known as the BMS symmetries. Similar to the two dimensional conformal symmetry, the generators of BMS$_3$ and BMS$_4$ are not globally well-defined and divided into two parts: super-translation  and super-rotation.  They are the infinite extensions of translation and rotation of the Poincare symmetry.
 
In this paper we focus on Einstein gravity in four dimensions. The BMS$_4$ algebra is given by
\begin{eqnarray}\label{BMS4-algebra}
 \nonumber [L_m,L_n]&=&(m-n)L_{m+n},\qquad [\bar L_m,\bar L_n]=(m-n)\bar L_{m+n},\qquad [L_m,\bar L_n]=0,\\
  {[}L_l,M_{m,n}{]}&=&\({l+1\over2}-m\)M_{m+l,n},\qquad   {[}\bar L_l,M_{m,n}{]}= \({l+1\over2}-n\)M_{m,n+l}. 
 \end{eqnarray}
 where $l$, $m$ and $n$ are integers and Poincare sub-algebra is given  by $\{L_0, L_{\pm 1}, \bar L_0, \bar L_{\pm 1}, M_{00}, M_{01}, M_{10},$  $M_{11}\}$. The first six generators are the generators of the Lorentz symmetry and the last four  generate translation. Thus, $L_n$ and $\bar L_n$ for all integers $n$ are known as super-rotations and $M_{nm}$ are super-translations.
 
 A useful representation for the generators of   algebra \eqref{BMS4-algebra}  is achieved  if one starts  from a generic solution of Einstein gravity, in an appropriate gauge (known as BMS gauge), and imposes particular boundary conditions \cite{Barnich:4dBMS, aspects}. In the coordinate $x^0=u$, $x^1=r$, $x^2=\theta$ and $x^3=\phi$,  metric of the four-dimensional  asymptotically flat spacetime is given by
 \begin{equation}\label{4d metric}
 ds^2=e^{2\beta}{V\over r} du^2-2e^{2\beta} du dr+g_{AB}\left(dx^A-U^A du\right)\left(dx^B-U^B du\right)
 \end{equation} 
 where $A,B=2,3$ and $V,\beta,U^A, g_{AB}$ are functions of coordinates. The boundary conditions which result in \eqref{BMS4-algebra} are given by \cite{Barnich:4dBMS, aspects}
 \begin{align}\label{boundary condition}
 \nonumber g_{uu}&=-2r\partial_u\varphi-e^{-2\varphi}+ \bar\nabla\varphi +\mathcal{O}(r^{-1}),\qquad g_{ur}=-1+\mathcal{O}(r^{-2}),\\
g_{uA}&=\mathcal{O}(1),\qquad g_{rr}=g_{rA}=0,\qquad g_{AB}=r^2 \bar\gamma_{AB}+\mathcal{O}(r), 
 \end{align}
 where $\varphi=\varphi(u,\theta,\phi)$, $\bar\gamma_{AB}dx^Adx^B= e^{2\varphi}\left(d\theta^2+\sin^2\theta d\phi^2\right)$ and $\bar\nabla$ denotes the Laplacian with respect to $\bar \gamma_{AB}$.
 
 According to the  AdS/CFT correspondence, the symmetry of the dual field theory is  the same as the asymptotic symmetry of the gravity solutions. Thus, it makes sense to think of BMS symmetry as the symmetry of the dual theory of the asymptotically flat spacetimes. More precisely, algebra \eqref{BMS4-algebra} denotes the symmetries of the dual three-dimensional  theory  of the four-dimensional asymptotically flat spacetimes \eqref{4d metric}. The  observation of \cite{Bagchi:2012cy} is that the whole BMS$_3$ and the global part of BMS$_4$ algebra \eqref{BMS4-algebra} are given by an ultra-relativistic contraction of the conformal symmetry in  two and three-dimensions, respectively. Thus BMSFTs are ultra-relativistic field  theories.

 Asymptotically flat spacetimes given by \eqref{4d metric} are dual to some states in the dual BMSFT. In the gravity side, distinct solutions are characterized by functions $V,\beta,U^A$ and $g_{AB}$  satisfying some equations which are resulted in from the Einstein field equations. Although explicit forms of these functions are not known but it is possible to find their expansions in terms of the $r$ coordinate \cite{Barnich:4dBMS, aspects}. In this paper we restrict our  analysis to a specific class of \eqref{4d metric} with $U^A=0$. More explicitly, we consider asymptotically flat spacetimes with line-element
 \begin{equation}\label{RT metric}
  ds^2=-qdu^2-2 du dr+2 r^2 e^{P(u,\zeta,\bar \zeta)}d\zeta d\bar\zeta,
  \end{equation} 
  where $\zeta$ and $\bar \zeta$ are complex coordinates given by $\zeta=e^{i\phi}\cot {\theta\over 2}$ and  $\bar \zeta=e^{-i\phi}\cot {\theta\over 2}$ and $q$ is given by 
  \begin{equation}
  q=-{2 M(u)\over r}+r \partial_u P(u,\zeta,\bar\zeta)-e^{-P(u,\zeta,\bar \zeta)}\partial_\zeta\partial_{\bar\zeta} P.
  \end{equation}
  The functions $ M $ and $P$ must satisfy the following equation:
  \begin{equation}
  3M\partial_u P+2\partial_u M+e^{-P}\partial_\zeta\partial_{\bar\zeta}\left(e^{-p}\partial_\zeta\partial_{\bar\zeta}P\right)=0.
  \end{equation}
   The line-element \eqref{RT metric} is known as the Robinson-Trautman geometry (RT). The Schwarzschild black hole and  asymptotically flat time dependent solutions  can be put  in this form. 
  
  Starting from \eqref{4d metric} , one can find the generic form of the asymptotic killing vectors $\xi^\mu$ which respect the boundary conditions \eqref{boundary condition} \cite{Barnich:4dBMS, aspects}:
  \begin{align}\label{explicit form of AKV}
 \nonumber \xi^u&= f(u,\zeta,\bar \zeta),\\
  \nonumber \xi^\zeta&= F(\zeta)-{1\over r}e^{-P}\partial_{\bar\zeta}f,\\
  \nonumber \xi^{\bar\zeta}&= \bar F(\bar\zeta)-{1\over r}e^{-P}\partial_{\zeta}f,\\
  \xi^r&=-{1\over 2}r\left(f\partial_u P+k\right)+e^{-P}\partial_\zeta\partial_{\bar\zeta} f,
  \end{align}
  where
  \begin{equation}
  k=e^{-P}\left(\partial_\zeta\left(e^P F\right)+\partial_{\bar\zeta}\left(e^P \bar F\right)\right),
  \end{equation}
  \begin{equation}
  f=e^{P\over2}\left(T(\zeta,\bar\zeta)+{1\over 2}\int du\,e^{-{P\over 2}}k\right).
  \end{equation}
  $F(\zeta)$, $\bar F(\bar\zeta)$ and $T(\zeta,\bar\zeta)$ are arbitrary functions of the complex coordinates $(\zeta,\bar\zeta)$. Generators $L_n$, $\bar L_n$ and $M_{nm}$ of BMS$_4$ algebra \eqref{BMS4-algebra} are given by 
  \begin{align}\label{basic definition}
  \nonumber L_n&=\xi\left(F(\zeta)=-\zeta^{n+1},\, \bar F(\bar\zeta)=0,\, T(\zeta,\bar \zeta)=0\right),\\
  \nonumber \bar L_n&=\xi\left(F(\zeta)=0,\, \bar F(\bar\zeta)=-\bar\zeta^{n+1},\, T(\zeta,\bar \zeta)=0\right),\\
   M_{mn}&=\xi\left(F(\zeta)=0,\, \bar F(\bar\zeta)=0,\, T(\zeta,\bar \zeta)=\zeta^{m}\bar\zeta^n\right).
  \end{align}
  
 The RT spacetimes \eqref{RT metric} are characterized by two functions $M$ and $P$. Variation of these functions under the action of the asymptotic killing vectors \eqref{explicit form of AKV} reads 
  \begin{align}\label{variation of functions}
  \nonumber \delta_\xi P&=0, \\
  \delta_\xi M&=2f\partial_u M+3M\left(k+f\partial_u P\right)-e^{-P}\left[\partial_\zeta f\,\partial_{\bar\zeta}\left(e^{-P}\partial_\zeta\partial_{\bar\zeta}P\right)+\partial_{\bar\zeta} f\,\partial_{\zeta}\left(e^{-P}\partial_\zeta\partial_{\bar\zeta}P\right)\right].
  \end{align}
 It is clear from the variation of $M$ that the  transformation \eqref{explicit form of AKV} do not respect the  form of the RT solutions and after the BMS  transformation, the final spacetime is in the form of \eqref{4d metric} with $U^A\neq 0$. 
 \section{Anisotropic conformal boundary and Robinson-Trautman stress tensor}
 A possible way to study the flat-space holography is taking the flat-space limit from the AdS/CFT calculations. 
 In this method, it is necessary to find a corresponding asymptotically AdS metric  for each of the asymptotically flat spacetimes which are related by taking the flat-space limit. An interesting property of the RT solutions is that for all of the possible values of the cosmological constant (including zero value),  metric is given by \eqref{RT metric}. The functions   $q$ and $P$  of the asymptotically flat solutions are given by taking the flat-space limit from their asymptotically AdS counterparts. For the asymptotically AdS solutions, $q$ is given by 
 \begin{equation}\label{q for AdS}
  q=-{2 {M}(u)\over r}+r \partial_u P(u,\zeta,\bar\zeta)-e^{-P(u,\zeta,\bar \zeta)}\partial_\zeta\partial_{\bar\zeta} P+ {r^2\over\ell^2}
  \end{equation}
where $\ell$ is the AdS radius.

 In this section, starting from the  asymptotically AdS RT solutions, we take appropriate flat-space limit from the components of the stress tensor  to propose asymptotically flat  RT stress tensors.  Then we justify our proposal.  

We consider Einstein gravity with negative cosmological constant in four-dimensions,
\begin{equation}\label{action0}
S_0={1\over 16 \pi G}\int dx^4\sqrt{-g}\left(R+{6\over\ell^2}\right).
\end{equation}

RT spacetimes, with $q$ given by \eqref{q for AdS} and appropriate functions $P$ satisfying the equations of motion, are solutions of this theory. The quasi-local stress tensor of these solutions is given by the Brown and York method \cite{Brown.AdS, Brown.qe},
 \begin{equation}\label{def.BY}
    T^{\mu\nu}={2\over\sqrt{-\gamma}}\dfrac{\delta S}{\delta \gamma_{\mu\nu}}, 
    \end{equation}  
 where $\gamma_{\mu\nu}=g_{\mu\nu}-n_\mu n_\nu$ is the boundary metric and $n_\mu$ is the outward pointing normal vector
to the boundary. $S$ is given by
\begin{equation}
S=S_0-{1\over 8\pi G}\int_{\partial M} d^3x {\mathcal{K}}+S_{ct},
\end{equation}
 where  $\mathcal{K}$  is  the extrinsic curvature of the boundary and $S_{ct}$ has the following form
 \begin{equation}\label{CTerm}
 S_{ct}=-{1\over 4 \pi G \ell}\int_{\partial M} d^3 x \sqrt{-\gamma}\left(1-{\ell^2\over 4}R_{(3)}\right),
 \end{equation}
where $R_{(3)}$ is the Ricci scalar of $\gamma_{\mu\nu}$. $S_{ct}$ has to be  added to remove the divergent terms at the boundary \cite{Brown.AdS}. Using \eqref{def.BY}-\eqref{CTerm} we find  the stress-tensor components as 
 \begin{align}\label{stress tensor AdS}
 \nonumber T_{uu}&={M\over4\pi G r \ell },\\
 \nonumber T_{u\zeta}&=T_{\zeta u}=-{\ell\over16\pi G r}\partial_{\zeta}\left(e^{-P}\partial_\zeta\partial_{\bar\zeta}P\right),\\
  \nonumber T_{u\bar\zeta}&=T_{\bar\zeta u}=-{\ell\over16\pi G r}\partial_{\bar\zeta}\left(e^{-P}\partial_\zeta\partial_{\bar\zeta}P\right),\\
   \nonumber T_{\zeta\zeta}&=-{\ell^3 e^{P}\over16\pi G r}\partial_{\zeta}\left(e^{-P}\partial_u\partial_{\zeta}P\right),\\
   \nonumber T_{\bar\zeta\bar\zeta}&=-{\ell^3 e^{P}\over16\pi G r}\partial_{\bar\zeta}\left(e^{-P}\partial_u\partial_{\bar\zeta}P\right),\\
   T_{\zeta\bar\zeta}&=T_{\bar\zeta\zeta}={\ell Me^{P}\over 8\pi G r}
 \end{align}
 We would like to take the flat-space limit from these components and propose a stress tensor for the asymptotically flat RT solution. The flat-space limit is given by $G/\ell^2\to 0$ while keeping $G$ fixed\footnote{$G/\ell^2$ is dimensionless in four-dimensions. }. It is clear from \eqref{stress tensor AdS} that in this limit the  stress tensor components are either zero or singular. Thus the direct limit is not well-defined. However, it is possible to define the stress tensors of the asymptotically flat cases by taking limit from the multiplication of  corresponding components in the asymptotically AdS cases to some powers of $G/\ell^2\to 0$. This method has been used previously in \cite{ Fareghbal:2013ifa,Fareghbal.Hosseni,Fareghbal:2016hqr,   Fareghbal.A.O}, for several cases. We expect that this method results in correct components of RT stress tensor. However, in all of the previous examples \cite{ Fareghbal:2013ifa,Fareghbal.Hosseni,Fareghbal:2016hqr,   Fareghbal.A.O}, the stress tensor components were just a function of $\ell$ or $1/\ell$.  $\ell^3$ factors in $T_{\zeta\zeta}$  and $T_{\bar\zeta\bar\zeta}$ are new. Apparently, the method of  \cite{ Fareghbal:2013ifa,Fareghbal.Hosseni,Fareghbal:2016hqr,   Fareghbal.A.O} needs to be refined in this case. In this paper we do not intend to study this problem but we would like to restrict our calculation to the case that  the method of  \cite{ Fareghbal:2013ifa,Fareghbal.Hosseni,Fareghbal:2016hqr,   Fareghbal.A.O} is applicable. Thus we consider the RT solutions whose  $P$ function is u-independent. Although this choice of  $P$ reduces the generality of our results but it still consists a large class of asymptotically flat solutions including Schwarzschild black hole.
 
 Another point is that the conservation of stress tensor of the asymptotically flat case requires that the components with one  $u$ index, become non-symmetric. The fact that BMSFTs are not relativistic theories may justify this non-symmetric stress tensors.  From the bulk  point of view although BMS symmetry is asymptotic symmetry which includes four-dimensional Poincare sub-group but boundary BMSFT$_3$ does not contain three-dimensional Poincare symmetry.  
 
 Putting  all together and setting $G=1$,  we propose the following components for the asymptotically flat RT solutions with $u$-independent $P$:
  \begin{align}\label{stress tensor flat}
 \nonumber T_{uu}&={M\over4\pi  r  },\\
 \nonumber T_{u\zeta}&=-{1\over16\pi  r}\partial_{\zeta}\left(e^{-P}\partial_\zeta\partial_{\bar\zeta}P\right),\qquad \,\,\,\,\,T_{\zeta u}=0,\\
  \nonumber T_{u\bar\zeta}&=-{1\over16\pi  r}\partial_{\bar\zeta}\left(e^{-P}\partial_\zeta\partial_{\bar\zeta}P\right),\,\,\,\,\,\qquad T_{\bar\zeta u}=0,\\
   \nonumber T_{\zeta\zeta}&=0,\qquad\qquad\qquad\qquad\qquad\qquad T_{\bar\zeta\bar\zeta}=0,\\
      T_{\zeta\bar\zeta}&=T_{\bar\zeta\zeta}={ Me^{P}\over 8\pi  r},
 \end{align}
 where $M=M(u)$  and $P=P(\zeta,\bar\zeta)$, using equation of motion, satisfy 
 \begin{equation}\label{EOM flat and u-independent p}
  2\partial_u M+e^{-P}\partial_\zeta\partial_{\bar\zeta}\left(e^{-p}\partial_\zeta\partial_{\bar\zeta}P\right)=0.
  \end{equation}
  From the above equation we conclude that $M$ is a linear function of $u$ coordinate. In the rest of this section we try to justify our proposal \eqref{stress tensor flat}.

Now we want to show that the proposed RT stress tensor \eqref{stress tensor flat} results in the correct charges of the symmetry generators. To do so, we use the Flat/BMSFT correspondence which proposes that the stress tensor \eqref{stress tensor flat}, calculated in the gravity side, is also the stress tensor of BMSFT. $u$, $\zeta$ and $\bar\zeta$ are  coordinates of the spacetime on which BMSFT lives  and $r$ is just a constant in this view.  The main challenging problem is to determine  the spacetime metric on which BMSFT lives. For the asymptotically AdS RT metric, given by \eqref{RT metric} and \eqref{q for AdS}, the dual CFT lives on the spacetime whose metric is simply given by the standard definition of the conformal boundary. For the current case,  metric of the boundary reads
\begin{equation}\label{metric of boundary AdS}
 ds^2={r^2\over\ell^2}\left(-du^2+2\ell^2 e^{P} d\zeta d\bar\zeta\right).
 \end{equation} 
Definition of the conformal infinity for the asymptotically flat spacetimes leads to a two-dimensional metric which is not suitable for  our analysis. The proposal to deal with this problem is to use anisotropic conformal infinity as the spacetime  where BMSFT  lives on it. This proposal has been succefully used  in  \cite{ Fareghbal:2013ifa,Fareghbal.Hosseni,Fareghbal:2016hqr,   Fareghbal.A.O} to compute the charges using flat-space stress tensor. The main idea goes back to \cite{Horava:2009vy} in which the authors have studied anisotropic conformal infinity in the context of holography for the non-relativistic theories.  The fact that BMSFTs have ultra-relativistic conformal symmetry is a  hint  that the methods of relativistic theories does not work. Currently we do  not know a well-established procedure to define the anisotropic conformal infinity in the context of the flat-space holography. However, in all of the previous works \cite{ Fareghbal:2013ifa,Fareghbal.Hosseni,Fareghbal:2016hqr,  Fareghbal.A.O} the geometry of   spacetime on which the BMSFT lives, was the same as the parent CFT whose contraction resulted in BMSFT. The only difference is that the AdS radius $\ell$ must be replaced by an appropriate power of Newton's constant $G$ which has the same dimension as $\ell$. If we accept the lessons of \cite{ Fareghbal:2013ifa,Fareghbal.Hosseni,Fareghbal:2016hqr,   Fareghbal.A.O}, the dual BMSFT  of RT solution lives on a spacetime with the following metric,
\begin{equation}\label{metric of boundary flat}
 ds^2={r^2\over G}\left(-du^2+2 G e^{P} d\zeta d\bar\zeta\right).
 \end{equation}
For $G=1$ and assuming $P$ as an $u$-independent function, we can use \eqref{stress tensor flat} and \eqref{metric of boundary flat} to calculate the symmetry charges. Before doing this calculation and using the fact that $M$ depends on $u$,  it is not difficult to show that the conservation equation  of stress tensor, $\nabla_\mu T^{\mu\nu}=0$, leads to  \eqref{EOM flat and u-independent p} which is a result of equations of motion in the gravity side. This is a good check for the correctness of our assumptions.

To calculate the  charges $Q_\xi$, associated to a symmetry generator $\xi$, we use the Brown and York formula \cite{Brown.qe},
\begin{equation}\label{def-of-charge}
Q_\xi=\int_\Sigma d\sigma \sqrt{\sigma}v^\mu\xi^\nu T_{\mu\nu},
\end{equation}
 where $\Sigma$ is a $u$-constant surface in the anisotropic conformal boundary, $\sigma_{ab}$ is the metric of $\Sigma$ and $v^\mu$ is the unit timelike vector normal to $\Sigma$. Using \eqref{explicit form of AKV}, \eqref{stress tensor flat} and \eqref{metric of boundary flat} we find,
 \begin{equation}\label{final charge}
 Q_\xi=\frac{1}{16 \pi}\int d\zeta d\bar\zeta e^{P}\left[4fM-\left(F\partial_\zeta+\bar F\partial_{\bar\zeta}\right)\left(e^{-P}\partial_\zeta\partial_{\bar\zeta}P\right)\right].
 \end{equation}
 This  is exactly the same as the results of the covariant phase space method in \cite{Barnich:2011mi}.

 \section{Correlators of stress tensor}
 We can use \eqref{stress tensor flat} and \eqref{variation of functions} to find the correlation functions of stress tensor. The procedure is  similar  to the CFT case. We assume that the correlators of BMSFT stress tensor are invariant under the action of the global part of the BMSFT symmetry. Using the asymptotic symmetry generators \eqref{explicit form of AKV} and the definition of BMS symmetry generators  \eqref{basic definition}, the global part is given by $\{L_0, L_{\pm 1}, \bar L_0, \bar L_{\pm 1}, M_{00}, M_{01}, M_{10},$  $M_{11}\}$ for fixed and large values of $r$ coordinate. 
 
 According to \eqref{stress tensor flat}, the stress tensor components are given in terms of $M$ and $P$ functions. Thus, in order to vary correlators we need to know the variation of $M$ and $P$ under the action of the global part of the BMS symmetry. This is given by   \eqref{variation of functions}. Since $\delta_\xi P=0$, all of the correlators consisting just $P$ functions are zero. This shows that all of the  correlators which do not have at least one $T_{uu}$ or $T_{\zeta\bar\zeta}$  are zero. Moreover, $\delta_\xi P=0$ expresses that to study asymptotically flat spacetimes we can restrict ourselves to the class of solutions with  particular $P$ functions.  The simplest case is $P=0$. As a result, using \eqref{stress tensor flat} we see that all of the correlators can be determined by using, $\{\langle M^1\rangle, \langle M^1 M^2\rangle, \langle M^1 M^2 M^3\rangle,\cdots \}$ where the index in  $M^i$ denotes the $i$-th insertion corresponding to the point  $(u_i,\zeta_i,\bar\zeta_i)$. Invariance of the correlators under the action of the global part results in
 \begin{equation}\label{1point M}
    \Phi_1:= \langle M^1\rangle=0
     \end{equation}    
 \begin{equation}\label{2pointM}
 \Phi_2:=\langle M^1 M^2\rangle={C\ \over (\zeta_1-\zeta_2)^3(\bar\zeta_1-\bar\zeta_2)^3}
 \end{equation}
 \begin{equation}\label{3point function}
\Phi_3:= \langle M^1 M^2 M^3\rangle= {\bar C\over (\zeta_1-\zeta_2)^{3\over 2}(\zeta_1-\zeta_3)^{3\over 2} (\zeta_2-\zeta_3)^{3\over 2} (\bar\zeta_1-\bar\zeta_2)^\frac32 (\bar\zeta_1-\bar\zeta_3)^\frac32 (\bar\zeta_2-\bar\zeta_3)^\frac32}
 \end{equation}
 where $C$ and $\bar C$ are constants. 
 Using \eqref{1point M}-\eqref{3point function} we find that  the non-zero  two and three point functions  are  $\langle T_{uu}^1 T_{uu}^2 \rangle, \langle T_{uu}^1 T_{\zeta\bar\zeta}^2  \rangle, \langle T_{\zeta\bar\zeta}^1 T_{\zeta\bar\zeta}^2 \rangle, \langle T_{uu}^1 T_{uu}^2 T_{uu}^3\rangle, \langle T_{uu}^1 T_{uu}^2 T_{\zeta\bar\zeta}^3  \rangle, \langle T_{uu}^1 T_{\zeta\bar\zeta}^2 T_{\zeta\bar\zeta}^3\rangle, \langle T_{\zeta\bar\zeta}^1 T_{\zeta\bar\zeta}^2 T_{\zeta\bar\zeta}^3  \rangle   $. The two point functions are proportional to $\Phi_2$ and  the three point functions are proportional to $\Phi_3$. Furthermore, the correlators are independent of  u coordinate. Our results are in complete agreement with the results of \cite{Bagchi:2016bcd} where the correlators of the primary operators of BMSFT  have been calculated by considering highest-weight representation of the BMS$_4$ algebra. Comparing our results with \cite{Bagchi:2016bcd} shows that in the case of the stress tensor operators, the eigenvalues of $L_0$ and $\bar L_0$ are  $3/2$. $L_0+\bar L_0$ is identified as the dilatation operator and $L_0-\bar L_0$ is the rotation operator. Thus we find that the weight under the dilatation operator is $3$. This may be an interesting properties of the three dimensional BMSFTs and needs more studies.

\section{Conclusion} 
 In this paper we generalized the method of \cite{ Fareghbal:2013ifa,Fareghbal.Hosseni,Fareghbal:2016hqr,   Fareghbal.A.O} to  four-dimensional asymptotically flat RT solutions. On the gravity side the quasi-local stress tensor yields the correct charges of the BMS symmetry generators. On the field theory side, the structure of correlation functions are consistent with the previous work \cite{Bagchi:2016bcd}. This shows that when an asymptotically flat metric is given by  taking the flat-space limit from an asymptotically AdS spacetime, one can use the dictionary of AdS/CFT to gain some insights into flat space holography.  
 This paper is the first step to study BMSFT$_3$ by using the method of \cite{ Fareghbal:2013ifa,Fareghbal.Hosseni,Fareghbal:2016hqr,   Fareghbal.A.O}. The final goal is to consider a generic four-dimensional asymptotically flat metric which satisfies the BMS boundary conditions and find quasi-local stress tensor  correlators.
 
 In all of our calculations in this paper and \cite{ Fareghbal:2013ifa,Fareghbal.Hosseni,Fareghbal:2016hqr,   Fareghbal.A.O} it is assumed that BMSFTs live on the anisotropic  conformal infinity of the asymptotically flat spacetimes. Its metric can be found  when the asymptotically flat geometry is given by taking the flat-space limit of an asymptotically AdS spacetime. However, it is still an open question to develop a systematic procedure to find the anisotropic conformal infinity in the context of flat-space holography.

\subsubsection*{Acknowledgements}
The authors would like to thank Mohammad Asadi, Seyed Morteza Hosseini and Pedram Karimi  for  useful comments.  This work is supported by  Iran National Science Foundation (INSF), project No. 9526713.

\appendix


\end{document}